# Selective Excitation of Coupled Resonators via Complex Frequency Driving: Enhanced Efficiency and Crosstalk Suppression

Deepanshu Trivedi[1], Laraib Niaz[1], Andrea Alù[3,4], and Alex Krasnok[1,2]

[1]*Department of Electrical Engineering, Florida International University, 33174, Miami, USA*

[2]*Knight Foundation School of Computing and Information Sciences, Florida International University, 33174, Miami, USA*

[3]*Photonics Initiative, Advanced Science Research Center, City University of New York, New York, NY 10031, USA*

[4]*Physics Program, Graduate Center, City University of New York, New York, NY 10016, USA*

*To whom correspondence should be addressed:* akrasnok@fiu.edu

## Abstract

Controlling individual elements of coupled resonator systems poses a significant challenge, as conventional real-frequency pulses suffer from inefficiency and crosstalk, limiting fidelity and scalability. To address this challenge, we propose and explore the use of complex frequency excitations, tailoring the driving signal waveform to match the target complex reflection zeros. We demonstrate that complex frequency driving can achieve near-unity selected energy storage efficiency ($\eta \approx 100\,\%$) in a single resonator, substantially exceeding the performance of optimized Gaussian pulses ($\eta_{\mathrm{max}} \approx 80\,\%$). In a coupled three-resonator system, our method yields significantly higher efficiency ($\eta \approx 92-95\,\%$) along with vastly improved selectivity and crosstalk suppression compared to conventional Gaussian pulse excitations of the same duration. Our technique achieves dynamic critical coupling, providing a powerful paradigm for high-fidelity, selective control, crucial for advancing scalable complex systems for sensing and computing.

**Keywords:** Complex frequency excitations, Coupled resonators, Microwave circuits, Selective addressing, Crosstalk suppression, Excitation efficiency.

## 1. Introduction

Coupled resonant systems are ubiquitous building blocks in modern physics and engineering, underpinning technologies ranging from optical filters and multiplexers (*1*, *2*) to microwave circuits for communications (*3*) and sensing (*4*), and forming the backbone of many quantum computing architectures, particularly in circuit quantum electrodynamics (cQED) (*5–8*). In these systems, multiple resonant elements (e.g., optical cavities, microwave resonators, qubits) interact with each other, often mediated through a shared communication bus, such as a waveguide or a transmission line (schematically depicted in **Figure 1**). While this coupling enables interaction and information exchange, it also introduces significant challenges for selective control. Excitation signals intended for one resonator can inadvertently affect its neighbors, leading to crosstalk, reduced fidelity of operations, and limitations on the density and scalability of integrated devices.

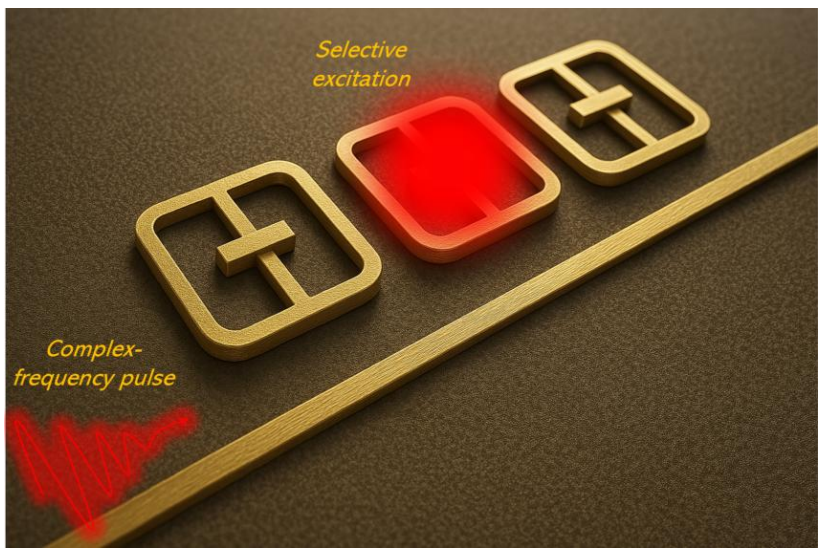

*Figure 1. Schematic illustration of multiple resonators coupled to a common microwave waveguide. This shared communication line facilitates interactions but also allows for crosstalk, complicating the selective control of individual resonators. Complex-frequency excitation offers a pathway to address a target resonator (e.g., the central one) with high fidelity while minimizing unwanted excitation (disturbance) of neighboring resonators, overcoming limitations associated with traditional real-frequency driving schemes.*

Traditional methods for exciting resonators typically employ real-frequency driving signals, often shaped as Gaussian or other smoothly varying pulses, centered at the resonance frequency of the target element (*9, 10*). While intuitive, this approach faces fundamental limitations. Firstly, impedance mismatch between the driving source/waveguide and the resonator leads to inherent reflections, limiting the maximum achievable energy transfer efficiency (*11*). Secondly, when resonators are closely spaced spectrally – a common scenario in densely integrated systems or systems with tunable elements – the finite spectral bandwidth of real-frequency pulses inevitably leads to off-resonant excitation of neighboring elements, causing significant crosstalk (*12*).

Various pulse shaping techniques have been developed to mitigate these issues, such as DRAG (Derivative Removal by Adiabatic Gate) pulses in quantum control (*9, 13*), but these often require complex calibration and may not fully suppress leakage, especially in strongly coupled or disordered systems.

Complex frequency excitations offer a promising route to overcome these limitations (*14–16*). Instead of driving the system at a real frequency, corresponding to the resonance of the desired element, in this approach we utilize an excitation signal that oscillates at a complex-valued frequency, specifically chosen to engage with the complex zero in the system's reflection coefficient. These reflection zeros, located at $\omega_z$ in the complex frequency plane, are associated with the response of the resonator-waveguide system, and related to its poles ($\omega_p$) which dictate the natural oscillation frequencies and decay rates (*17, 18*). Driving the system at a frequency $\omega_z$ located in the appropriate half of the complex frequency plane allows for perfect impedance matching at the port, effectively creating a time-dependent "critical coupling" scenario where the incoming energy is fully stored in the system without reflection (*19–23*). This requires an excitation signal with a precisely tailored, exponentially growing amplitude that counteracts the system's intrinsic decay rate.

This approach shares foundational principles with concepts like coherent perfect absorption (CPA) and time-reversed lasing (*15, 24*), where specific real frequencies associated with system zeros are targeted to achieve complete absorption or emission control. However, while CPA often focuses on achieving zero scattering in steady-state scenarios, typically using multiple coherent input beams, here we focus on dynamic, single-port excitation of specific resonant modes within a coupled system for the purpose of high-fidelity state preparation and energy storage. The emphasis is on leveraging reflection zeros at complex-valued frequencies ($\Gamma(\omega_z) = 0$) for efficient and selective energy delivery into individual components of a potentially complex network, a distinct application focus crucial for quantum control and signal routing.

In the following we present a theoretical and numerical investigation into the efficacy of complex frequency excitation for controlling coupled microwave resonator systems, comparing it directly with conventional Gaussian pulse driving of the same resonator network. We begin by establishing the foundational circuit model for a single resonator coupled to a waveguide, analyzing its scattering properties and defining the concept of excitation efficiency. Using this model, we demonstrate the superior efficiency achievable with complex frequency driving compared to optimized Gaussian pulses, validating our simulations against analytical predictions from Temporal Coupled Mode Theory (TCMT) (*18*, *21*, *25*, *26*), see Supplementary Materials for more details. We then extend our analysis to a more complex and practically relevant scenario: a system of three distinct resonators coupled to a common feedline. This multi-resonator system allows us to rigorously assess the selectivity of complex frequency excitation in the presence of potential crosstalk, employing quantitative metrics for performance evaluation. In our work, we focus on parameters commonly associated with coplanar waveguide resonators used in circuit quantum electrodynamics (*8*). We utilize sophisticated simulation tools, including the Quantum Circuit Analyzer Tool (QuCAT) (*27*, *28*) for eigenmode analysis and the PathWave Advanced Design System (ADS) for time-domain transient simulations, complemented by analytical calculations using the ABCD matrix formalism in MATLAB. Through detailed comparisons of transient responses and energy distribution among the resonators under both Gaussian and complex frequency excitation schemes, we quantify the significant advantages offered by the complex frequency approach in terms of both energy storage efficiency and, crucially, suppression of unwanted excitation in neighboring resonators. Our findings demonstrate that complex frequency driving provides a robust and highly effective method to achieve high-fidelity, selective control in complex multi-resonator systems, paving the way for advancements in the development of scalable quantum information processors.

The paper is structured as follows: the next section details the theoretical framework, including the circuit model for the resonator-waveguide system, the definition of complex poles and zeros, the formulation of excitation signals, and quantitative performance metrics. Then, we present the results, starting with the single-resonator case, followed by an in-depth analysis of the three-resonator system, comparing efficiency and selectivity for both excitation methods. Finally, we provide an expanded discussion of the implications of these findings, including physical mechanisms, the impact of realistic imperfections like loss and nonlinearity, practical implementation challenges, scalability and comparison with other control methods.

## 2. Results and Analysis

We start our analysis with the canonical model of a single resonant cavity coupled to a feedline or waveguide [Figure 2(a)]. We assume that the resonator is described by an intrinsic resonance frequency determined by its internal structure, modeled here by a parallel inductor-capacitor (LC) circuit [Figure 2(b)] with frequency $\omega_0 = 1/\sqrt{LC}$, where L and C are the inductance and capacitance. The resonator is coupled to a single-mode waveguide (modeled as a transmission-line with characteristic impedance $Z_0$, typically 50 Ω) via a coupling element, represented here by a coupling capacitance $C_\kappa$. This coupling allows energy to be exchanged between the waveguide and the resonator, and it also provides a channel for the stored energy in the resonator to decay back into the waveguide. This decay occurs at the external decay rate, $\gamma_{ext}$, which is determined by the coupling strength.

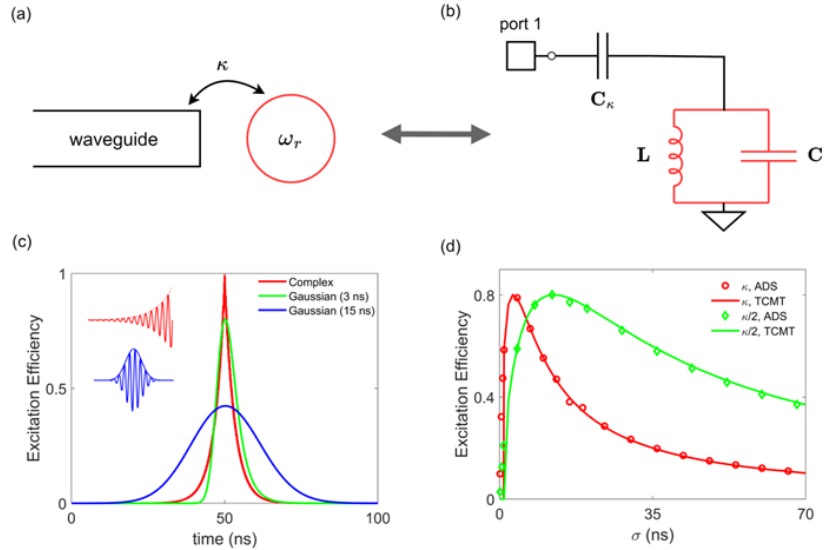

**Figure 2.** (a) Schematic representation of a resonant cavity with eigenfrequency $\omega_r$ coupled via $\gamma_{ext}$ to a feeding waveguide. (b) Equivalent lumped-element circuit model. The waveguide is modeled by the input port with impedance $Z_0$, coupled capacitively ($C_\kappa$) to a parallel LC resonator. (c) Simulated excitation efficiency $\eta$ versus time for a single resonator using: a long Gaussian pulse ($\sigma = 15$ ns, blue), a short Gaussian pulse ($\sigma = 3$ ns, green), and optimal excitation signal oscillating at a complex frequency (red). (d) Peak excitation efficiency versus Gaussian pulse width $\sigma$ for two different external coupling strengths ($\gamma_{ext}$ corresponding to $C_\kappa = 50$ fF, and $\gamma_{ext}/2$ corresponding to smaller $C_\kappa$). Circles and diamonds represent numerical simulations (ADS), while solid lines show analytical predictions from TCMT.

In the case of capacitive coupling $C_\kappa$ to a high-quality factor resonator, the external decay rate is related to the external quality factor $Q_{ext}$ by $\gamma_{ext} = \dfrac{\omega_r}{2Q_{ext}}$, where $\omega_r$ is the resonance frequency of the *coupled* system (slightly shifted from $\omega_0$ due to the coupling) and $Q_{ext}$ is inversely proportional to the square of the coupling capacitance for weak coupling (*8*, *11*), $Q_{ext} \approx \frac{C}{\omega_r Z_0 C_\kappa^2}$.
Thus, a stronger coupling (larger $C_\kappa$) leads to a lower $Q_{ext}$ and a higher external decay rate $\gamma_{ext} \propto C_\kappa^2$. In the Hamiltonian formalism, as in conventional TCMT, a coupling constant $\kappa$ can be used, which is related to the decay rate as $\kappa = \sqrt{2\gamma_{ext}}$. This parameter $\kappa$ represents the amplitude coupling rate between the resonator and the waveguide.

The interaction with the waveguide is defined by the system's scattering parameters, particularly the reflection coefficient $\Gamma(\omega)$, defined for a load impedance $Z_L(\omega)$ describing by the coupled resonator as $\Gamma(\omega) = \frac{Z_L(\omega) - Z_0}{Z_L(\omega) + Z_0}$. Here, $Z_L(\omega)$ is the frequency-dependent impedance of the coupled resonator system. Analyzing $\Gamma(\omega)$ in the complex frequency plane ($\omega = \omega_{re} + j\omega_{im}$) reveals crucial information encoded in its poles and zeros (*17*, *18*). The poles, denoted by $\omega_p$, are the complex frequencies for which the reflection diverges ($\Gamma(\omega_p) \to \infty$), corresponding to the natural eigenfrequencies of the coupled system. For a stable passive system, these poles reside in the upper half of the complex frequency plane ($\text{Im}(\omega_p) > 0$). The real part, $\omega_r = \text{Re}(\omega_p)$, indicates the resonant frequency of the coupled system, while the imaginary part determines the total decay rate, $\gamma_{tot} = |\text{Im}(\omega_p)|$. In a system with only external coupling considered, this total decay rate equals the external decay rate, $\gamma_{tot} = \gamma_{ext}$. Conversely, the zeros, denoted by $\omega_z$, are the complex frequencies where the reflection vanishes ($\Gamma(\omega_z) = 0$). This condition implies perfect impedance matching ($Z_L(\omega_z) = Z_0$), allowing an appropriately shaped incident wave to be perfectly absorbed without reflection. In lossless systems, poles and zeros appear in complex conjugate pairs. The complex reflection zeros $\omega_z$ are the targets of the complex frequency excitation scheme investigated here. It is also important to consider the residue, or strength, associated with each pole-zero pair, which influences how strongly the mode interacts with the input port at real frequencies. A small residue indicates weak coupling or a faint resonance signature in reflection measurements near the real frequency axis, making efficient excitation with conventional real-frequency pulses challenging.

To evaluate the effectiveness of different excitation strategies, we define the excitation efficiency $\eta(t)$ as the ratio of energy stored in the resonator mode $a(t)$ to the total injected energy supplied by the excitation signal $s_{ex}(t)$ up to time $t$:

$$\eta(t) = \frac{|a(t)|^2}{\int_0^t |s_{ex}(t')|^2 \, dt'} * 100\%. \tag{1}$$

We then compare two distinct excitation approaches: first, we will use conventional Gaussian pulses, which employ a standard real-frequency pulse shape, centered near the target resonance frequency $\text{Re}(\omega_z) \approx \omega_r$. The mathematical form is given by, $s_{ex}(t) = \frac{A}{\sqrt{2\pi}\sigma} \exp\left(-\frac{(t-t_0)^2}{2\sigma^2}\right) \exp(j\text{Re}(\omega_z)t)$. Here, A represents the amplitude, $\sigma$ is a parameter controlling the pulse duration, and $t_0$ is the time of the pulse center. The efficiency of excitation with Gaussian pulses oscillating at real frequencies is limited by reflections arising from impedance mismatch, as the reflection coefficient is generally non-zero at real frequencies, $\Gamma(\text{Re}(\omega_z)) \neq 0$. The optimal choice of pulse width $\sigma$ to maximize efficiency typically corresponds to matching the pulse bandwidth to the resonator linewidth, i.e., $\sigma \sim 1/2\gamma_{ext}$. Second, we will explore complex frequency excitations: this strategy directly targets a specific reflection zero $\omega_z = \text{Re}(\omega_z) + j\text{Im}(\omega_z)$, where $\text{Im}(\omega_z) < 0$ for zeros corresponding to stable poles (decaying modes). The required excitation waveform takes the form, $S_{ex}(t) = A\exp[j\text{Re}(\omega_z)t]\exp[-\text{Im}(\omega_z)t]$. A key feature of this waveform is their exponential growth, governed by the positive exponent $-\text{Im}(\omega_z) > 0$. This growth is precisely tailored to dynamically compensate for the resonator's energy decay rate $\gamma_{ext}$, which, assuming negligible intrinsic loss, is approximately equal to $|\text{Im}(\omega_z)|$. By counteracting the energy loss in real time, the driving signal maintains the condition $\Gamma \approx 0$, effectively achieving dynamic critical coupling and enabling the excitation efficiency $\eta$ to approach unity.

We first simulate the single resonator system (**Figure 2**) using $L = 2$ nH, $C = 0.4$ pF ($\omega_0/2\pi \approx 5.63$ GHz), $Z_0 = 50$ $\Omega$, and $C_\kappa = 50$ fF. From the optimal Gaussian pulse width observed in simulations ($\sigma_{opt} \approx 3$ ns, see **Figure 2d**), we estimate the corresponding external decay rate as $\gamma_{ext} \approx 1/2\sigma_{opt} \approx 1.67 \times 10^8$ rad/s (or $\gamma_{ext}/2\pi \approx 27$ MHz). This corresponds to an external Q-factor $Q_{ext} = \omega_r/(2\gamma_{ext}) \approx 106$. The detailed analytical derivations for the excitation efficiency under both complex frequency and Gaussian driving schemes, supported by numerical validation (Figure S1), are provided in the *Supplementary Materials*.

**Figure 2(c)** compares the temporal excitation efficiency $\eta(t)$ for different drives: a long Gaussian pulse ($\sigma = 15$ ns), significantly longer than the resonator's decay time ($1/2\gamma_{ext}$), achieves a peak efficiency of approximately $\eta_{peak} \approx 40\%$, primarily due to substantial energy loss through reflections. An optimized short Gaussian pulse, with a duration matched to the resonator's decay time ($\sigma = 3$ ns $\approx 1/2\gamma_{ext}$), performs considerably better, reaching a peak efficiency $\eta_{peak} \approx 80\%$. This value represents the practical upper limit for this system when using conventional Gaussian driving, fundamentally constrained by the inherent impedance mismatch at real frequencies. In stark contrast, complex frequency excitations, precisely targeting the calculated reflection zero $\omega_z$, achieves near-perfect energy absorption, with efficiency $\eta$ rapidly approaching unity after a brief initial transient phase.

**Figure 2(d)** further reinforces this conclusion by showing the peak efficiency achieved with Gaussian pulses as a function of the pulse width $\sigma$ for two different coupling strengths. The simulation results obtained using PathWave ADS (represented by points) clearly show the existence of an optimal $\sigma$ for each coupling strength, yielding a maximum efficiency around 80%. These numerical findings are in excellent agreement with analytical predictions derived independently from Temporal Coupled Mode Theory (TCMT) (*18*, *21*, *29*) (solid lines), providing strong validation for both the lumped-element circuit model and the theoretical framework employed. This baseline single-resonator analysis unequivocally establishes the fundamental efficiency advantage offered by complex frequency driving ($\eta = 100\%$) compared to optimally tuned real-frequency driving ($\eta_{max} < 100\%$).

To investigate excitation performance in a more complex environment, and showcase the opportunity of selective excitation, we analyze a system comprising three distinct resonators ($i = 1,2,3$) coupled to a common feedline via individual capacitances $C_{ci}$. The feedline itself is connected to the external port via a primary coupling capacitance $C_\kappa$ (**Figure 3(a)**). The specific parameters used in our simulations are $L_1 = L_2 = L_3 = 2$ nH, $C_1 = C_2 = C_3 = 0.4$ pF, $C_{c1} = 10$ fF, $C_{c2} = 11$ fF, $C_{c3} = 12.1$ fF, and $C_\kappa = 10$ pF. These parameters define the individual uncoupled frequencies $\omega_{0i} = 1/\sqrt{L_i C_i}$ and determine their external coupling rates $\gamma_{ext,i}$ which scale approximately as $C_{ci}^2$. The interaction between the resonators, mediated by the shared feedline, leads to hybridization of the individual modes into system eigenmodes with distinct eigenfrequencies. These eigenfrequencies and their dependence on system parameters (e.g., varying $C_1$) are determined using the Quantum Circuit Analyzer Tool (QuCAT)(28), which operates by diagonalizing the circuit Hamiltonian derived from the lumped-element model. **Figure 3(b)** illustrates the eigenfrequency dispersion diagram, showing the evolution of the three system eigenmodes as $C_1$ is varied. The observed avoided crossings are characteristic signatures of mode interactions and energy exchange between coupled resonant modes.

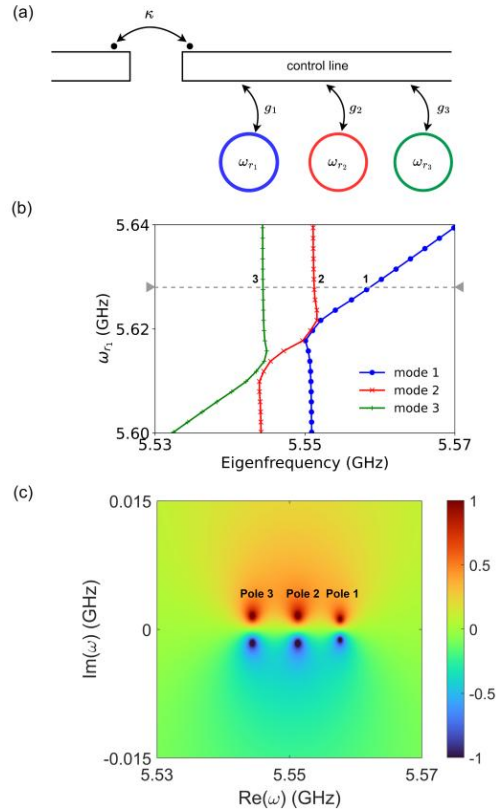

*Figure 3. (a): Schematic of the three-resonator system. Each LC resonator is coupled ($C_{c1}$, $C_{c2}$, $C_{c3}$) to a common control line, which couples ($C_\kappa$) to the external port ($Z_0$). (b): Eigenfrequency dispersion calculated using QuCAT, showing the real parts of the three system eigenmodes versus $C_1$. Avoided crossings indicate interactions. (c): Complex poles and zeros of the reflection coefficient $\Gamma(\omega)$ calculated via ABCD matrix method for a specific $C_1$ (grey dashed line in (b)).*

Complementary to the eigenfrequency analysis, the scattering properties of the system are analyzed by calculating the complex poles ($\omega_{pi}$) and zeros ($\omega_{zi}$) of the reflection coefficient $\Gamma(\omega)$. This calculation is performed numerically using the ABCD matrix method (11, 30) implemented in MATLAB. The method involves constructing the ABCD matrix for each circuit element (resonators represented by their impedance, coupling capacitors, and feedline segments) and then cascading these matrices to obtain the overall ABCD matrix for the entire network as seen from the input port terminated by the characteristic impedance $Z_0$. From the overall matrix, $\Gamma(\omega)$ can be derived, and its poles and zeros found numerically. The poles $\omega_{pi}$ correspond to the complex eigenfrequencies of the system, with $\text{Re}(\omega_{pi})$ matching the eigenfrequencies found by QuCAT and $\text{Im}(\omega_{pi})$ giving the total decay rate $\gamma_{tot,i}$ of each mode. The zeros $\omega_{zi}$ satisfy the condition $\Gamma(\omega_{zi}) = 0$ and represent the frequencies required for perfect absorption via complex frequency driving. These numerically calculated complex zeros $\omega_{zi}$ are then utilized to define the precise frequency and exponential growth rate for the complex frequency excitation sources within the PathWave ADS time-domain simulations. **Figure 3(c)** shows the calculated poles and zeros in the complex frequency plane for a specific system configuration, illustrating their characteristic pairing for a reciprocal system.

To quantitatively assess performance in the multi-resonator system, we introduce two specific metrics evaluated at the end of the excitation pulse (time $T$). The first metric is *Target Selectivity* ($\mathcal{S}_i^{target}$), defined as the fraction of energy stored in the intended target resonator $i$ relative to the total energy stored across all resonators:

$$\mathcal{S}_i^{target} = \frac{\eta_i(T)}{\sum_{j=1}^{3} \eta_j(T)} . \tag{2}$$

A value of $\mathcal{S}_i^{target}$ close to 1 indicates that the excitation successfully directed most of the stored energy into the desired resonator. The second metric is the *Crosstalk Suppression Ratio* ($\mathcal{C}_i$), defined as the ratio of the energy stored in the target resonator $i$ to the maximum energy stored in any *other* non-target resonator $j \neq i$:

$$\mathcal{C}i = \frac{\eta_i(T)}{\max_{j \neq i}(\eta_j(T))} . \tag{3}$$

A higher value of $\mathcal{C}_i$ signifies better suppression of unwanted energy transfer, indicating superior isolation of the target resonator from its neighbors during excitation. These metrics provide a quantitative basis for comparing the selectivity performance of different driving protocols.

We examine the performance of conventional Gaussian pulses, centered at the real part of the reflection zero associated with each target resonator, $\text{Re}(\omega_{zi})$, for $i = 1,2,3$. We use Gaussian pulse durations consistent with those used in the time domain simulations for the complex frequency pulse excitation discussed below. While a different choice of $\sigma$ or more sophisticated pulse shaping (like DRAG) might marginally improve selectivity for Gaussian driving, the results presented here using a standard Gaussian form are representative of the inherent difficulties this approach faces due to fixed real-frequency operation, namely impedance mismatch reflections and spectral overlap causing crosstalk. The specific Gaussian pulse duration used for each resonators are: $\sigma_1 = 30$ ns, $\sigma_2 = 22$ ns, $\sigma_3 = 22$ ns. The Gaussian pulse is defined with a pulse width (standard deviation) $\sigma_i$, but the effective pulse width (measured between points where the amplitude drops to $1/\sqrt{2}$ of the peak) is approximately $1.62\sigma_i$. However, with an optimally chosen pulse i.e., a long duration pulse ($\sigma_{opt} \approx 2.5\sigma_i$), the same system can achieve efficiency as high as 86%.

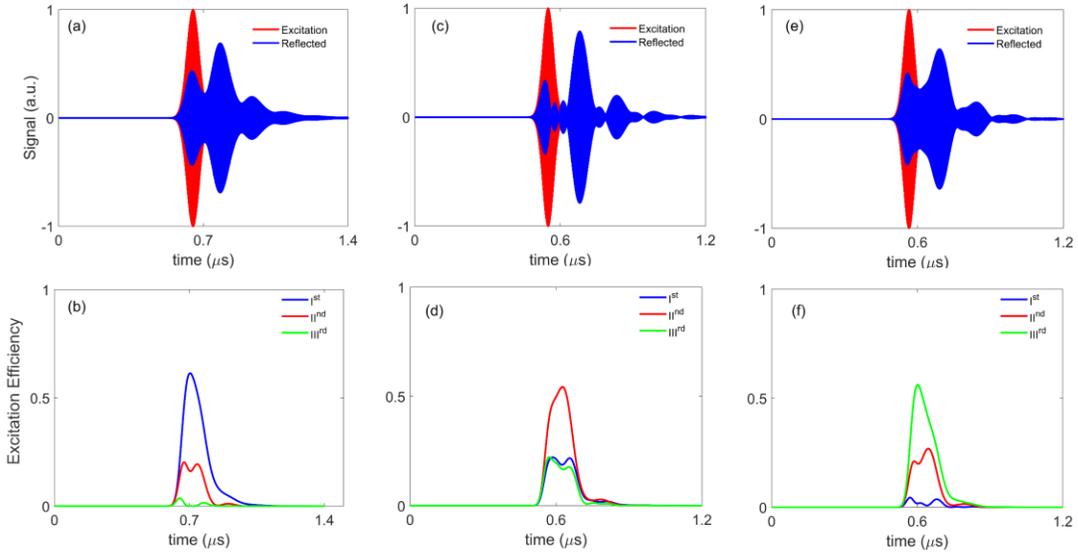

*Figure 4.* Transient analysis of the three-resonator system under Gaussian pulse excitation ($\sigma_{1,2,3} = 30, 22, 22$ ns) (a, c, e): Incident pulse (red, scaled) and reflected signal (blue) when targeting (a) resonator 1 ($\text{Re}(\omega_{z1})$), (c) resonator 2 ($\text{Re}(\omega_{z2})$), and (e) resonator 3 ($\text{Re}(\omega_{z3})$). (b, d, f): Corresponding time-dependent excitation efficiencies ($\eta_i$) within each resonator: $\eta_1$ (blue), $\eta_2$ (red), $\eta_3$ (green).

When targeting resonator 3 (the most strongly coupled via $C_{c3} = 12.1$ fF) using a Gaussian pulse centered at $\text{Re}(\omega_{z3})$, the simulation shows significant reflection (**Figure 4(e)**). The energy distribution analysis (**Figure 4(f)**) indicates that resonator 3 captures most of the stored energy, reaching a peak efficiency of $\eta_3 \approx 56\%$. Leakage into the other resonators depends upon the proximity of their corresponding modes. Since mode corresponding to resonator 2 is closely spaced, the leakage into it is relatively high at $\eta_2 \approx 20\%$, while the estimated leakage into resonator 1 is $\eta_1 \approx 2\%$. Calculating the selectivity metrics defined earlier, we find a total stored efficiency $\eta_{tot} \approx 78\%$, a target selectivity $\mathcal{S}_3^{target} = 0.709$, and a crosstalk suppression ratio $\mathcal{C}_3 = 2.665$. While overall efficiency is reasonable, the selectivity is poor.

However, a similar situation occurs while exciting resonator 1 (the most weakly coupled via $C_{c1} = 10$ fF) using a Gaussian pulse centered at $\text{Re}(\omega_{z1})$: the simulation shows significant reflection similar to the first case [**Figure 4(a)**]. The energy distribution analysis [**Figure 4(b)**] indicates that resonator 1 captures the majority of the stored energy, reaching a peak efficiency $\eta_1 \approx 61\%$. Due to the closer spectral proximity of resonator 2 compared to resonator 3, there is notable leakage into the corresponding mode of resonator 2, estimated at $\eta_2 \approx 16\%$ while leakage into resonator 3 remains negligible at $\eta_3 \approx 0.1\%$. Calculating the selectivity metrics defined earlier, we find a total stored efficiency $\eta_{tot} \approx 78\%$, a target selectivity $\mathcal{S}_1^{target} = 0.772$, and a crosstalk suppression ratio $\mathcal{C}_1 = 3.405$. Similar to resonator 3, the overall efficiency is limited, although the selectivity is slightly improved in this case. Targeting resonator 2 ($C_{c2} = 11$ fF), which exhibits intermediate coupling strength and lies spectrally between the closely spaced modes of resonators 1 and 3, results in significant leakage from these neighboring modes, thereby reducing the overall excitation efficiency of the system. **Figure 4(c)** shows reflections, and **Figure 4(d)** indicates a peak efficiency in resonator 2 $\eta_2 \approx 54\%$. Due to its central position, resonator 2 exhibits symmetric leakage into resonator 1 and 3, with each receiving approximately $\eta_1 \approx 19\%$ and $\eta_3 \approx 17\%$ The selectivity metrics reflect this scenario: total stored energy efficiency $\eta_{tot} \approx 73\%$, target selectivity $\mathcal{S}_2^{target} = 0.508$, and crosstalk suppression ratio $\mathcal{C}_2 = 2.05$. In this case, selectivity is the lowest due to significant leakage into adjacent modes. The excitation efficiency is also limited to 54%, the lowest among all three resonators cases. Overall, these results demonstrate that Gaussian pulse excitation suffers not only from limited efficiency due to reflections, but also it exhibits significant vulnerability to crosstalk and energy coupling, particularly when dealing with spectrally dense systems.

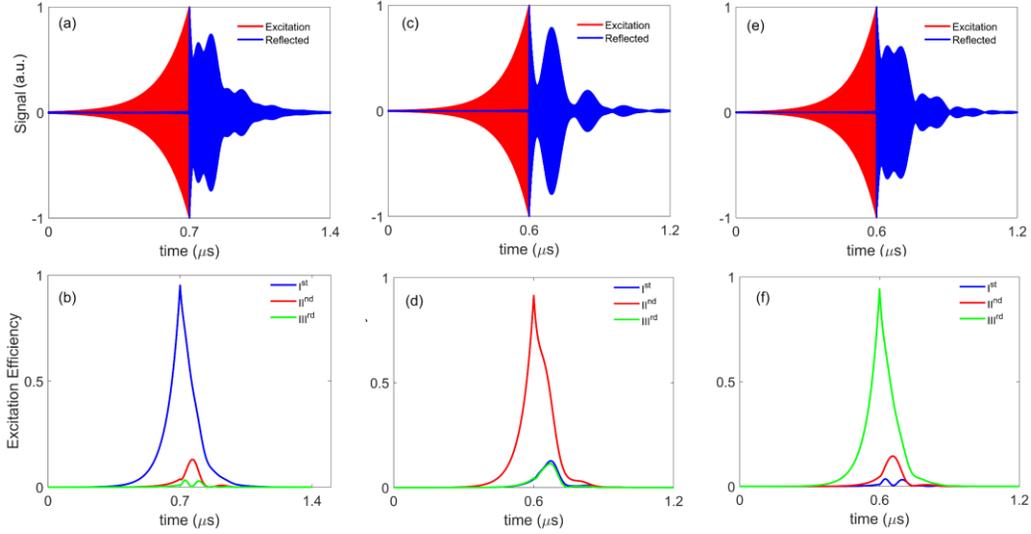

*Figure 5.* Transient analysis of the three-resonator system under complex frequency excitation. (a, c, e): Incident complex frequency signal (red, exponentially growing) and reflected signal (blue) when exciting at (a) $\omega_{z1}$, (c) $\omega_{z2}$, and (e) $\omega_{z3}$. Note the near-zero reflection during excitation. (b, d, f): Corresponding time-dependent excitation efficiencies ($\eta_i$) within each resonator: $\eta_1$ (blue), $\eta_2$ (red), $\eta_3$ (green).

We now evaluate the performance of complex frequency excitation signals, specifically tailored to target the calculated complex zeros $\omega_{zi}$ for each resonator. When targeting resonator 3 using the complex frequency $\omega_{z3}$, the simulation (**Figure 5(e)**) shows the characteristic features of this method: an exponentially growing incident signal (red envelope) and, crucially, a near-zero reflected signal (blue) during the entire excitation phase. This confirms the establishment of dynamic critical coupling, allowing for efficient energy transfer. The energy distribution analysis (**Figure 5(f)**) reveals a high peak efficiency in the target resonator, $\eta_3 \approx 94\%$. However, some leakage into the adjacent resonator 2 is observed ($\eta_2 \approx 5\%$), while leakage into resonator 1 remains minimal ($\eta_1 \approx 1\%$). The selectivity metrics are: total stored $\eta_{tot} \approx 100\%$ (reflecting near-perfect absorption), target selectivity $\mathcal{S}_3^{target} = 0.943$, and crosstalk suppression ratio $\mathcal{C}_3 = 20$. Compared to Gaussian driving for this target, the efficiency is significantly higher (94% vs 56%), the target selectivity is improved (0.943 vs 0.709), and the crosstalk suppression is better (20 vs 2.665).

Similarly, when targeting the weakly coupled resonator 1 using the complex frequency $\omega_{z1}$. The transient response [**Figure 5(a)**] again shows negligible reflection during excitation. The energy distribution [**Figure 5(b)**] demonstrates highly efficient and selective excitation of this mode: the efficiency in resonator 1 reaches $\eta_1 \approx 95\%$. Leakage occurs primarily into the adjacent resonator 2 ($\eta_2 \approx 4\%$) with minimal energy in resonator 3 ($\eta_3 \approx 1\%$). The selectivity metrics quantify this impressive response: total stored $\eta_{tot} \approx 100\%$, target selectivity $\mathcal{S}_1^{target} = 0.95$, and crosstalk suppression ratio $\mathcal{C}_1 \approx 24.92$. Comparing these values to the Gaussian case ($\eta_1 \approx 61\%$, $\mathcal{S}_1 \approx 0.772$, $\mathcal{C}_1 \approx 3.405$) highlights the excellent improvement in both efficiency and selectivity offered by complex frequency driving.

Finally, targeting resonator 2 with the complex frequency $\omega_{z2}$ yields similar results, as expected given its intermediate coupling and relative spectral isolation. **Figure 5(c)** shows reflectionless energy transfer. The energy analysis [**Figure 5(d)**] confirms that essentially maximum absorbed energy is stored in resonator 2, with $\eta_2 \approx 92\%$. However, there is symmetrical leakage into resonators 1 and 3 estimated at $\eta_1, \eta_3 \approx 4\%$. Thee performance metric for this case: total stored $\eta_{tot} \approx 100\%$, target selectivity $\mathcal{S}_2^{target} \approx 0.915$, and crosstalk suppression ratio $\mathcal{C}_2 \approx 20.33$. This represents a substantial enhancement over the best Gaussian case ($\eta_2 \approx 54\%$, $\mathcal{S}_2 \approx 0.508$, $\mathcal{C}_2 \approx 2.05$). The results from the three-resonator system, quantitatively assessed using efficiency and selectivity metrics, provide compelling evidence for the superiority of complex frequency excitation. It achieves maximum efficiency $\approx 95\%$ while simultaneously offering significantly enhanced selectivity, particularly overcoming the limitations of conventional methods when addressing specific modes within a complex, coupled environment, such as spectrally close modes. This performance arises because, at these complex frequencies, the eigenmodes corresponding to the three target resonances become highly localized within their respective resonators, minimizing modal overlap and suppressing undesired coupling. A complete quantitative comparison of the target selectivity, crosstalk suppression ratio, and excitation efficiency for both driving protocols is detailed in the *Supplementary Materials* (Tables S1-S6).

## 3. Discussion

The presented analysis highlights the superiority of complex frequency excitations for high-fidelity control of coupled resonators. The method fundamentally overcomes the impedance mismatch limitation of real-frequency pulses by dynamically achieving critical coupling through waveform shaping matched to the system's reflection zeros $\omega_z$. This approach enables very high energy transfer efficiency (92-95 %), offering a clear advantage over Gaussian pulse excitation. While an optimized Gaussian pulse can achieve efficiencies as high as 86%, it does so at the expense of speed. Additionally, long-duration pulses are generally undesirable in several research areas, including quantum computing and quantum electrodynamics, due to their limitations in temporal resolution and increased susceptibility to decoherence. This efficiency gain translates directly to reduced power requirements or faster operation times for achieving a desired energy state in the target resonator. In the multi-resonator context, the selectivity advantages demonstrated are arguably even more critical than the efficiency gains. The quantitative metrics ($S_i^{target}$, $C_i$) highlight the superior ability of complex frequency driving to isolate target resonators. Additionally, it efficiently excites the resonator 2, which typically suffers from high leakage into both adjacent modes, yet still achieves high excitation efficiency at ($\eta_2 \approx 92$ %) and good selectivity ($S_1 \approx 0.915$, $C_1 \approx 20.33$), whereas Gaussian driving performs poorly ($\eta_1 \approx 54$ %, $S_1 \approx 0.508$, $C_1 \approx 2.05$). This capability stems from the fact that complex frequency driving leverages the unique phase and amplitude dynamics dictated by the *entire* complex zero $\omega_z = \text{Re}(\omega_z) + j\text{Im}(\omega_z)$.

While the ideal theoretical performance is impressive, practical implementations must contend with system imperfections. Our analysis assumed lossless, linear resonators. Relaxing these assumptions reveals important considerations. Real resonators inevitably exhibit intrinsic losses due to factors like dielectric absorption, conductor resistance, or two-level system defects. These losses introduce an additional energy decay channel, characterized by an internal quality factor $Q_i$ and an internal decay rate $\gamma_{int} = \omega_r / (2Q_i)$. The total decay rate of the resonator mode then becomes the sum of external and internal rates, $\gamma_{tot} = \gamma_{ext} + \gamma_{int}$. This modification shifts the location of the system's pole in the complex frequency plane to $\omega_{p'} \approx \omega_r - j(\gamma_{ext} + \gamma_{int})$. Importantly, the reflection zero, which is the target for complex frequency driving, also shifts due to the presence of loss, moving to approximately $\omega_{z'} \approx \omega_r - j(\gamma_{ext} - \gamma_{int})$ (*14*). To maintain the zero-reflection condition, the complex frequency drive must now target this modified zero $\omega_{z'}$. If the external coupling remains stronger than the internal loss ($\gamma_{ext} > \gamma_{int}$), the zero $\omega_{z'}$ still resides in the lower half-plane, meaning perfect absorption ($\Gamma(\omega_{z'}) = 0$) is, in principle, still achievable with an appropriately adjusted exponentially growing waveform. However, the energy perfectly absorbed by the system is now dissipated through *both* the external coupling channel and the internal loss channel. Consequently, the maximum efficiency for storing energy *within the* *resonator itself* (relative to the total input energy) is no longer unity but becomes limited by the branching ratio of the decay rates. Specifically, the maximum achievable stored energy efficiency is given by $\eta_{max} \approx \gamma_{ext}/(\gamma_{ext}+\gamma_{int}) = Q_{tot}/Q_{ext}$, where $1/Q_{tot} = 1/Q_{ext} + 1/Q_i$. This underscores that even with perfect absorption enabled by complex frequency driving, achieving high *storage* efficiency still necessitates resonators with high intrinsic quality factors ($Q_i \gg Q_{ext}$).

Another critical factor, especially relevant in superconducting circuits often used for quantum information processing, is nonlinearity. At higher excitation powers, required for fast quantum gate operations, the Kerr effect can become significant. This effect causes an intensity-dependent shift in the resonator's frequency: $\omega_r \rightarrow \omega_r + K|a|^2$, where $K$ is the Kerr coefficient (related to the nonlinearity of the Josephson junction inductance or kinetic inductance) and $|a|^2$ represents the energy stored in the resonator (proportional to the average photon number). Since the locations of the system's poles ($\omega_p$) and zeros ($\omega_z$) depend on the resonance frequency $\omega_r$, the Kerr effect makes these critical frequencies intensity-dependent. As energy accumulates in the resonator during the excitation pulse, $\omega_r$ shifts, causing $\omega_p$ and $\omega_z$ to dynamically move in the complex plane. A complex frequency drive targeting a fixed, pre-calculated $\omega_z$ based on the low-power response will thus lose its perfect absorption property as the resonator energy increases. Maintaining high fidelity and efficiency under strong driving conditions would necessitate employing adaptive control waveforms where the instantaneous frequency and growth rate of the drive signal, $\omega(t)$, dynamically track the moving zero $\omega_z(t)$. Designing such adaptive waveforms presents a significant control theory challenge but may be essential for leveraging complex frequency driving in high-power applications like fast qubit gates (*31*).

Despite the dramatic improvement in selectivity, our simulations revealed that complex frequency driving does not completely eliminate crosstalk, particularly when targeting mode 2, where about 4% leakage into mode 1 and mode 3 was observed (**Fig. 5d**, $C_2 \approx 20.33$). Understanding the origin of this residual leakage is important. It occurs even though the reflection coefficient at the target frequency $\omega_{z2}$ is ideally driven to zero. The primary mechanism is likely related to the non-orthogonality of the system eigenmodes in the presence of coupling to the shared feedline. While the drive $S_{ex}(t)$ at frequency $\omega_{z2}$ is specifically tailored to perfectly cancel reflections associated with mode 2, the electromagnetic field generated by this drive still exists within the common waveguide. This driving field inevitably has a non-zero spatial overlap with the mode profiles of the other resonators (1 and 3). Even though the drive frequency is off-resonant for modes 1 and 3, this spatial overlap allows the drive to exert a small force on these modes, leading to some energy transfer. This effect is expected to be more pronounced for modes that are spectrally closer or share a stronger indirect coupling via the bus. This mechanism differs from the dominant source of crosstalk in Gaussian driving, which is the direct spectral overlap of the pulse's finite bandwidth with adjacent resonance frequencies. Secondary effects, such as transients during the finite ramp-up or ramp-down of the exponential waveform, or the influence of higher-order poles and zeros not considered in the simple single-pole approximation, could also contribute to the observed leakage. Further theoretical work might explore optimizing the pulse envelope beyond a pure exponential (e.g., using windowed exponentials or adding corrective tones) to specifically minimize the excitation of non-target modes by actively canceling the field overlap effects.

Translating these theoretical advantages into experimental practice presents several challenges. Firstly, it requires accurate knowledge of the target complex zeros $\omega_{zi}$ of the fabricated device. These can be determined through careful characterization using, for example, vector network analyzer (VNA) measurements of the reflection coefficient $\Gamma(\omega)$ followed by numerical fitting or analytical continuation to find the zeros in the complex plane. However, these zeros can be sensitive to fabrication variations, temperature drifts, and changes in the electromagnetic environment, necessitating robust in-situ calibration procedures. Machine learning techniques or adaptive algorithms that iteratively refine the drive waveform based on measured system response could prove valuable (*32*). Secondly, generating the required exponentially growing waveforms with high fidelity demands arbitrary waveform generators (AWGs) possessing sufficient bandwidth (to accommodate the real frequency component $\text{Re}(\omega_z)$ and the fast rise time) and dynamic range (to accurately produce the exponential envelope over the desired pulse duration). The presence of noise in the control lines or the AWG output could also potentially degrade the perfection of the zero-reflection condition.

Regarding scalability, applying complex frequency driving to systems with many more coupled resonators ($N \gg 3$) appears feasible in principle. The main challenge lies in the increased complexity of the scattering response. An N-resonator system will generally possess N poles and N zeros. Calculating these accurately requires robust numerical methods (like extensions of the ABCD approach or other scattering matrix techniques). Furthermore, as the spectral density of modes increases, the zeros may become closely spaced, potentially requiring finer frequency resolution and longer pulse durations to maintain selectivity, which could conflict with the need for fast operations. However, the fundamental principle of targeting a specific zero associated with a target resonator remains valid, suggesting that complex frequency driving could still offer significant advantages over real-frequency methods in large-scale systems, provided accurate characterization and waveform generation are possible.

Finally, it is useful to consider complex frequency driving in the context of other advanced control methodologies. Techniques like optimal control theory (e.g., GRAPE algorithm) numerically design complex pulse shapes to achieve specific state transfers or unitary operations with high fidelity, often implicitly suppressing leakage and crosstalk. Shortcuts to adiabaticity (STA) aim to speed up adiabatic processes while suppressing unwanted transitions. Complex frequency driving offers a more analytical and physically intuitive approach compared to purely numerical optimal control, as it directly targets a specific physical property (the reflection zero). It might be less versatile than optimal control for implementing arbitrary unitary gates but could be significantly simpler to design and potentially more robust for state preparation tasks. It could also potentially be combined with other techniques; for instance, the exponential envelope could be modified based on STA principles or optimal control could be used to refine the waveform around the analytically derived complex frequency drive to further enhance performance in the presence of specific imperfections.

**Conclusion**

This work provided a comprehensive theoretical and numerical comparison between conventional Gaussian pulse excitation and complex frequency excitation for controlling single and multiple coupled microwave resonators connected via a common waveguide. Our analysis, rigorously validated through multiple simulation methods and employing quantitative performance metrics for efficiency and selectivity, unequivocally demonstrates the significant advantages of the complex frequency approach. By targeting the complex reflection zeros ($\omega_z$) of the system with exponentially growing waveforms, complex frequency driving overcomes the fundamental impedance mismatch limitations of real-frequency pulses, achieving near-unity energy storage efficiency ($\eta \approx 100\%$). More importantly, in a multi-resonator system prone to crosstalk, complex frequency excitation exhibits vastly superior selectivity ($S_i^{target} \to 1$) and crosstalk suppression ($C_i \gg 1$) compared to Gaussian pulses. It successfully excites target resonators even in the presence of closely spaced nearby modes, with high fidelity and significantly reduced energy leakage. This enhanced performance stems from the ability of complex frequency driving to dynamically achieve critical coupling and leverage the unique decay characteristics associated with each mode's complex zero. While practical implementation requires careful consideration of intrinsic losses, nonlinearities, calibration accuracy, and advanced waveform generation capabilities, the principles of complex frequency excitation offer a powerful and promising paradigm for high-fidelity, selective control in complex resonant systems. This approach leverages fundamental wave interference phenomena for precise control. This methodology holds significant potential for advancing applications in microwave engineering, photonics, and particularly for enabling scalable and robust quantum information processing platforms based on coupled resonator architectures.

# Supplementary materials

# Selective Excitation of Coupled Resonators via Complex Frequency Driving: Enhanced Efficiency and Crosstalk Suppression

Deepanshu Trivedi[1], Laraib Niaz[1], Andrea Alù[3,4], and Alex Krasnok[1,2]

[1]Department of Electrical Engineering, Florida International University, 33174, Miami, USA

[2]Knight Foundation School of Computing and Information Sciences, Florida International University, 33174, Miami, USA

[3]Photonics Initiative, Advanced Science Research Center, City University of New York, New York, NY 10031, USA

[4]Physics Program, Graduate Center, City University of New York, New York, NY 10016, USA

To whom correspondence should be addressed: akrasnok@fiu.edu

## S.1. Theoretical Derivation of Excitation Efficiency

This section provides a detailed analytical derivation of the excitation efficiency for a single-mode resonator coupled to a waveguide, as described by Temporal Coupled Mode Theory (TCMT). We analyze and contrast two distinct excitation schemes: a tailored complex frequency-matched pulse and a conventional Gaussian pulse.

### 1.1. Temporal Coupled Mode Theory Framework

The dynamics of the resonator's complex amplitude, denoted as $\tilde{a}(t)$, when driven by a single input port can be described by the following first-order differential equation. For this analysis, we neglect intrinsic resonator losses ($\gamma_{\text{int}} = 0$).

$$\frac{d\tilde{a}(t)}{dt} = -i\omega_0 \tilde{a}(t) - \gamma \tilde{a}(t) + \sqrt{2\gamma}\tilde{s}_{in}(t)$$

By analyzing the system in a reference frame rotating at the resonator's natural frequency $\omega_0$, the $i\omega_0 \tilde{a}(t)$ term is eliminated, simplifying the equation to:

$$\frac{d\tilde{a}(t)}{dt} = -\gamma \tilde{a}(t) + \sqrt{2\gamma}\tilde{s}_{in}(t) \quad \text{(S1)}$$

Here, $\gamma$ represents the coupling decay rate to the waveguide, and $\tilde{s}_{in}(t)$ is the complex envelope of the input signal, normalized such that $|\tilde{s}_{in}(t)|^2$ gives the input power.

The primary metric for evaluating the performance is the time-dependent **excitation efficiency** $\eta(t)$, defined as the ratio of the energy stored in the resonator to the total energy injected by the source up to time $t$:

$$\eta(t) = \frac{|\tilde{a}(t)|^2}{\int_0^t |\tilde{s}_{in}(\tau)|^2 d\tau} \quad (S2)$$

### 1.2. Impulse Response of the Resonator

To derive the system's response to an arbitrary input, we first determine its impulse response, $h(t)$, which serves as the Green's function. We consider an impulse excitation, $\tilde{s}_{in}(t) = \delta(t)$. The governing equation becomes:

$$\frac{d\tilde{a}(t)}{dt} = -\gamma \tilde{a}(t) + \sqrt{2\gamma} \delta(t) \quad (S3)$$

Integrating this equation across an infinitesimal window $[-\epsilon, \epsilon]$ around $t = 0$ establishes the initial condition for the resonator's amplitude immediately after the impulse:

$$\int_{-\epsilon}^{\epsilon} \left( \frac{d\tilde{a}(t)}{dt} + \gamma \tilde{a}(t) \right) dt = \sqrt{2\gamma} \int_{-\epsilon}^{\epsilon} \delta(t) dt \Rightarrow \tilde{a}(0^+) = \sqrt{2\gamma}$$

For $t > 0$, the system evolves according to the homogeneous equation, $\frac{d\tilde{a}(t)}{dt} = -\gamma \tilde{a}(t)$, yielding the solution $\tilde{a}(t) = \sqrt{2\gamma} e^{-\gamma t}$. The complete impulse response, enforcing causality with the Heaviside step function $\Theta(t)$, is:

$$h(t) = \sqrt{2\gamma} e^{-\gamma t} \Theta(t) \quad (S4)$$

The resonator's amplitude $\tilde{a}(t)$ for any arbitrary input signal $\tilde{s}_{in}(t)$ can then be found via the convolution integral:

$$\tilde{a}(t) = (h * \tilde{s}_{in})(t) = \int_0^t h(t - \tau) \tilde{s}_{in}(\tau) d\tau = \sqrt{2\gamma} \int_0^t e^{-\gamma(t-\tau)} \tilde{s}_{in}(\tau) d\tau$$

### 1.3. Case 1: Complex Frequency Excitation

For complex frequency excitation, the input signal is designed to counteract the resonator's natural decay, featuring an exponentially growing envelope:

$$\tilde{s}_{in}(t) = A_C e^{\gamma t} \Theta(t)$$

where $A_C$ is a constant amplitude.

- **Resonator Amplitude:** Substituting this signal into the convolution integral gives:

$$\tilde{a}(t) = \sqrt{2\gamma} A_C \int_0^t e^{-\gamma(t-\tau)} e^{\gamma\tau} d\tau = \sqrt{2\gamma} A_C e^{-\gamma t} \int_0^t e^{2\gamma\tau} d\tau \quad (S5)$$

Solving the integral yields:

$$\tilde{a}(t) = \sqrt{2\gamma} A_C e^{-\gamma t} \left[\frac{e^{2\gamma\tau}}{2\gamma}\right]_0^t = \frac{A_C}{\sqrt{2\gamma}} e^{-\gamma t}(e^{2\gamma t} - 1) = \frac{A_C}{\sqrt{2\gamma}}(e^{\gamma t} - e^{-\gamma t}) \quad (S6)$$

- **Energy Calculation:** The energy stored in the resonator, $E_{out}(t)$, is:

$$E_{out}(t) = |\tilde{a}(t)|^2 = \frac{A_C^2}{2\gamma}(e^{\gamma t} - e^{-\gamma t})^2 = \frac{A_C^2}{2\gamma}(e^{2\gamma t} - 2 + e^{-2\gamma t}) \quad (S7)$$

The total energy injected into the system, $E_{in}(t)$, is:

$$E_{in}(t) = \int_0^t |\tilde{s}_{in}(\tau)|^2 d\tau = A_C^2 \int_0^t e^{2\gamma\tau} d\tau = \frac{A_C^2}{2\gamma}(e^{2\gamma t} - 1) \quad (S8)$$

- **Excitation Efficiency:** The resulting efficiency $\eta(t)$ is the ratio of stored to input energy:

$$\eta(t) = \frac{E_{out}(t)}{E_{in}(t)} = \frac{\frac{A_C^2}{2\gamma}(e^{2\gamma t} - 2 + e^{-2\gamma t})}{\frac{A_C^2}{2\gamma}(e^{2\gamma t} - 1)} = \frac{(e^{\gamma t} - e^{-\gamma t})^2}{e^{2\gamma t} - 1} = \frac{(e^{2\gamma t} - 1)(1 - e^{-2\gamma t})}{e^{2\gamma t} - 1}$$

This simplifies to:

$$\eta(t) = 1 - e^{-2\gamma t} \quad (S9)$$

This result shows that $\eta(t)$ monotonically approaches unity as $t \to \infty$, demonstrating that perfect energy transfer is theoretically possible.

## 1.4. Case 2: Gaussian Pulse Excitation

Next, we analyze the response to a Gaussian input pulse centered at $t_0$ with a characteristic width $\sigma$:

$$\tilde{s}_{in}(t) = A_G e^{-\frac{(t-t_0)^2}{2\sigma^2}}$$

- **Resonator Amplitude:** The convolution integral for the Gaussian pulse is:

$$\tilde{a}(t) = A_G\sqrt{2\gamma}\int_{-\infty}^{t} e^{-\gamma(t-\tau)} e^{-\frac{(\tau-t_0)^2}{2\sigma^2}} d\tau = A_G\sqrt{2\gamma}e^{-\gamma t}\int_{-\infty}^{t} e^{\gamma\tau} e^{-\frac{(\tau-t_0)^2}{2\sigma^2}} d\tau \quad (S10)$$

By completing the square in the exponent of the integrand and integrating, the solution can be expressed in terms of the complementary error function, erfc($x$):

$$\tilde{a}(t) = A_G\sqrt{\pi\gamma}e^{-\gamma t+\gamma t_0+\frac{\gamma^2\sigma^2}{2}}\text{erfc}\left(\frac{\gamma\sigma^2 - t + t_0}{\sqrt{2}\sigma}\right)$$

The stored energy is therefore:

$$E_{out}(t) = |\tilde{a}(t)|^2 = A_G^2\pi\gamma e^{-2\gamma t+2\gamma t_0+\gamma^2\sigma^2}\left[\text{erfc}\left(\frac{\gamma\sigma^2 - t + t_0}{\sqrt{2}\sigma}\right)\right]^2 \quad (S11)$$

- **Energy Calculation:** The total energy of the input Gaussian pulse is constant for $t \to \infty$:

$$E_{in} = \int_{-\infty}^{\infty} |\tilde{s}_{in}(\tau)|^2 d\tau = A_G^2 \int_{-\infty}^{\infty} e^{-\frac{(\tau-t_0)^2}{\sigma^2}} d\tau = A_G^2 \sigma\sqrt{\pi} \quad (S12)$$

- **Excitation Efficiency:** The final efficiency for Gaussian excitation is:

$$\eta(t) = \frac{E_{out}(t)}{E_{in}} = \gamma\sqrt{\pi}e^{\gamma(\gamma\sigma^2-2t+2t_0)}\left[\text{erfc}\left(\frac{\gamma\sigma^2 - t + t_0}{\sqrt{2}\sigma}\right)\right]^2 \quad (S13)$$

This expression shows that the efficiency under Gaussian excitation is a more complex function of time and the pulse parameters $(\sigma, t_0)$. Unlike the complex frequency case, the temporal mismatch between the Gaussian shape and the resonator's exponential impulse response fundamentally limits the maximum achievable efficiency. As confirmed by numerical simulations and optimization, this maximum value is approximately **80%**, in stark contrast to the near-unity efficiency achieved with the temporally matched complex frequency pulse.

## 1.5. Numerical Verification of Excitation Efficiency

To visually validate the analytical results derived in the preceding sections, we performed numerical simulations of the excitation efficiency for both Gaussian and complex frequency driving schemes. The results are presented as density plots, which map the efficiency as a function of time and a key system parameter.

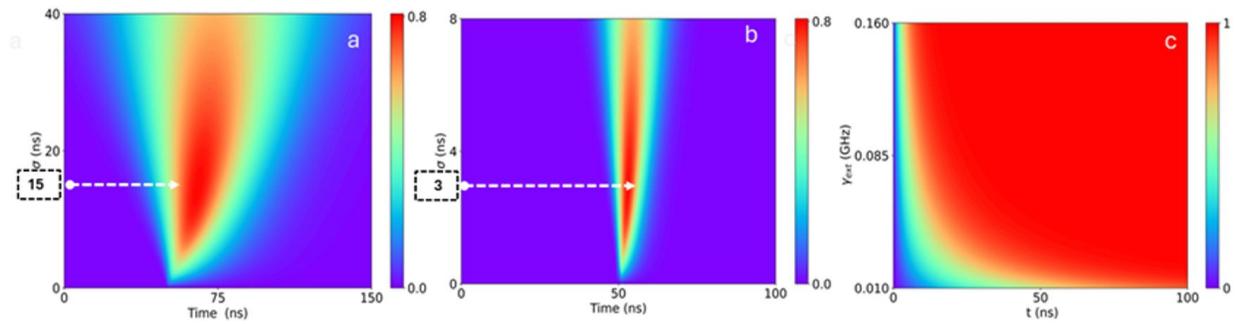

**Figure S1. Comparative Analysis of Excitation Efficiency.** The density plots illustrate the time-dependent excitation efficiency $\eta(t)$ for a single-port resonator system under different excitation protocols. The color scale indicates the efficiency, from 0 (purple) to maximum (red). (a), (b) Efficiency landscape for a Gaussian pulse excitation, plotted as a function of time and the pulse width $\sigma$ for two different coupling rates (see Figure 2 in the main text). The plot shows that the efficiency is highly dependent on both parameters, reaching a maximum value of approximately 80% for an optimal $\sigma$ at a specific point in time. (c) Efficiency landscape for the complex frequency-matched excitation, plotted as a function of time and the resonator's external coupling rate $\gamma_{ext}$. In stark contrast to the Gaussian case, the efficiency robustly and asymptotically approaches unity (100%) for all tested values of $\gamma_{ext}$, demonstrating perfect energy transfer over time.

The numerical results presented in **Figure S1** provide a compelling visual confirmation of the theoretical frameworks outlined in Equations (S9) and (S13).

**Gaussian Pulse Excitation (Fig. S1a and S1b):** The efficiency plots for the Gaussian pulse highlight a fundamental limitation of using pre-defined pulse shapes for resonator excitation. As shown in Equation (S13), the efficiency $\eta(t)$ is a complex function of the pulse width $\sigma$ and time $t$. Figure S1a and S1b numerically solve for this efficiency across a parameter sweep, demonstrating that an optimal efficiency is achieved only for a specific combination of $\sigma$ and observation time. The global maximum efficiency is visibly capped at approximately 80% (deep red color corresponds to 0.8 on the color bar). This limitation arises from the inherent temporal mismatch between the symmetric Gaussian shape of the input pulse and the asymmetric exponential decay of the resonator's impulse response ($h(t) \propto e^{-\gamma t}$). Because the pulse shape is not perfectly matched to the system's natural response, a portion of the input energy is always reflected, preventing complete energy absorption regardless of the pulse width or duration.

**Complex Frequency Excitation (Fig. S1c):** The results for complex frequency excitation stand in sharp contrast. The input signal, with its exponentially growing envelope ($\tilde{s}_{in}(t) \propto e^{\gamma t}$), is expressly designed to be the time-reversed counterpart of the system's impulse response. This temporal matching dynamically cancels reflections and ensures perfect impedance matching to the resonator. The resulting efficiency, derived as $\eta(t) = 1 - e^{-2\gamma t}$ in Equation (S9), is a function of time and the decay rate $\gamma$ but is not fundamentally limited by a shape mismatch. Figure S1c perfectly illustrates this principle. For any given coupling rate $\gamma_{ext}$ on the y-axis, the efficiency (color) rapidly increases with time along the x-axis and saturates at a value of 1 (100%). This demonstrates that, given sufficient time, the complex frequency protocol achieves near-perfect energy transfer into the resonator, a result that is robust across a wide range of system coupling parameters.

## 2. Detailed Efficiency and Selectivity Results

**Table S1:** *Target Selectivity under Gaussian Pulse Excitation.* Values indicate the fraction of total stored energy residing in each resonator (columns) when the system is excited with a Gaussian pulse centered at the real part of the reflection zero corresponding to the target indicated by the row label (Leftmost zero, Middle zero, Rightmost zero).

| Excitation at frequency of zero (Gaussian pulse) | Target Selectivity ($\mathcal{S}_i^{target}$) | | |
|---|---|---|---|
|  | Resonator 1 | Resonator 2 | Resonator 3 |
| Leftmost zero | 0.025 | 0.266 | 0.709 |
| Middle zero | 0.248 | 0.509 | 0.244 |
| Rightmost zero | 0.772 | 0.226 | 0.001 |

**Table S2:** *Target Selectivity under Complex Frequency Excitation.* Values indicate the fraction of total stored energy residing in each resonator (columns) when the system is excited with a complex frequency pulse matched to the reflection zero corresponding to the target indicated by the row label.

| Excitation at frequency of zero (Complex Frequency pulse) | Target Selectivity ($S_i^{target}$) | | |
|---|---|---|---|
| | Resonator 1 | Resonator 2 | Resonator 3 |
| Leftmost zero | 0.01 | 0.047 | 0.9428 |
| Middle zero | 0.04 | 0.915 | 0.045 |
| Rightmost zero | 0.952 | 0.038 | 0.010 |

**Table S3:** *Crosstalk Suppression Ratio under Gaussian Pulse Excitation*. Values indicate the ratio of energy in the target resonator to the maximum energy in any non-target resonator when the system is excited with a Gaussian pulse centered at the real part of the reflection zero corresponding to the target indicated by the row label.

| Excitation at frequency of zero (Gaussian pulse) | Crosstalk Suppression Ratio ($C_i$) | | |
|---|---|---|---|
| | Resonator 1 | Resonator 2 | Resonator 3 |
| Leftmost zero | 0.035 | 0.375 | 2.665 |
| Middle zero | 0.487 | 2.05 | 0.487 |
| Rightmost zero | 3.405 | 0.294 | 0.002 |

**Table S4:** *Crosstalk Suppression Ratio under Complex Frequency Excitation*. Values indicate the ratio of energy in the target resonator to the maximum energy in any non-target resonator when the system is excited with a complex frequency pulse matched to the reflection zero corresponding to the target indicated by the row label.

| Excitation at frequency of zero (Complex Frequency pulse) | Crosstalk Suppression Ratio ($C_i$) | | |
|---|---|---|---|
| | Resonator 1 | Resonator 2 | Resonator 3 |
| Leftmost zero | 0.010 | 0.05 | 20 |
| Middle zero | 0.044 | 20.33 | 0.049 |
| Rightmost zero | 24.92 | 0.040 | 0.010 |

**Table S5:** *Efficiency % under Complex Frequency Excitation*. Each row corresponds to targeting a specific resonator (Resonator 1, 2, or 3) using complex frequency excitation matched to its reflection zero. The columns show the percentage of the total input energy stored in each respective resonator ($\eta_1, \eta_2, \eta_3$) at the end of the pulse. Diagonal elements represent the efficiency in the target resonator.

| Excitation (Complex Frequency) | Efficiency (%) | | |
|---|---|---|---|
| | $\eta_1$ | $\eta_2$ | $\eta_3$ |
| Resonator 1 | 94.7 | 3.8 | 1 |
| Resonator 2 | 4 | 91.5 | 4.5 |
| Resonator 3 | 1 | 4.7 | 94 |

**Table S6:** *Efficiency % under Gaussian Pulse Excitation*. Each row corresponds to targeting a specific resonator (Resonator 1, 2, or 3) using Gaussian pulse excitation centered at the real part of its reflection zero. The columns show the percentage of the total input energy stored in each respective resonator ($\eta_1, \eta_2, \eta_3$) at the point of maximum total stored energy. Diagonal elements represent the efficiency in the target resonator.

| Excitation (Gaussian) | Efficiency (%) | | |
|---|---|---|---|
| | $\eta_1$ | $\eta_2$ | $\eta_3$ |
| Resonator 1 | 61.4 | 16.1 | 0.1 |
| Resonator 2 | 18.7 | 54.3 | 16.8 |
| Resonator 3 | 1.8 | 20 | 56.2 |